\documentclass[preprint,notfootinbib]{revtex4}
\usepackage{graphicx}
\graphicspath{{c:/user/dell/desktop/papermgb/plots}}
\usepackage{epsfig}
\usepackage{amssymb}
\usepackage{color}
\usepackage{amsmath}
\usepackage{lipsum}
\usepackage[font={small}]{caption}
\usepackage{graphics}

\usepackage{epsf}
\usepackage{psfrag}
\usepackage{epstopdf}
\usepackage{graphicx}
\usepackage{caption}
\usepackage{epstopdf}
\usepackage{epsfig}
\begin{document}
\title{ Observational Tests of Gauss-Bonnet Like Dark Energy Model}
\author{ Z. Molavi \footnote{Email:\text{zmolavi26@gmail.com}}~and~
A. Khodam-Mohammadi \footnote{Email:\text{khodam@basu.ac.ir
(corresponding author)}}}
 \affiliation{Department of
Physics, Faculty of Science, Bu-Ali Sina University, Hamedan 65178,
Iran}
\begin{abstract}
The consistency of some dynamical dark energy models based on
Gauss-Bonnet invariant, ${\cal G}$, is studiedcompared with
cosmological observational tests. The investigated models are
modified form of Gauss Bonnet dark energy, MGB-DE and two other
versions which are interacting MGB and $n_0$MGB. The energy density
of proposed models are combinations of powers of the Hubble rate, H,
and its time derivative. To inquire the performance of MGB dark
energy models, we have used data analyzing methods and numerical
solutions, in both background and perturbed levels, based on recent
observational data from SNIa, Baryon Acoustic Oscillations (BAO),
Hubble parameter, CMB data, and structure formation data surveys.
Employing joint data sets and comparing the results to those of
LCDM, show that all versions of MGB-DE predicts the expansion
history and evolution of structures appropriately as well as
$\Lambda$CDM. If we use pure late universe data set, we see that all
models of MGB-DE are successful in recent epoch, and there is not
any significant evidence against or in favor of $\Lambda$CDM,
whereas for early universe, statistical results indicate a
significantly better agreement for $\Lambda$CDM as compared to all
versions of MGB-DE models.
\end{abstract}
\maketitle
\section{Introduction}
\noindent Entering the era of precision cosmology, scientists faced huge amount of data,
received by several surveys from the mysterious sky. The accurate astrophysical data from
distant Ia supernovae \cite{Riess:1998cb}, \cite{Perlmutter:1997zf}, \cite{Hicken:2009dk},
cosmic microwave background anisotropy \cite{Komatsu:2008hk}, \cite{Larson:2010gs}, and large
scale galaxy surveys \cite{Abazajian:2004it}, \cite{Tegmark:2006az}, reveals that the
universe is nearly spatially flat and is definitely passing an accelerating expansion phase.
This is one of the most fundamental concepts in theoretical cosmology and particle physics.\\
During last decades, quite high number of models have been presented
in this context. These models are mainly categorized in two classes.
The first insists on modifying and extending the gravity itself,
named modified gravity. Modified gravity models assume that, the
present accelerating epoch is due to geometric effects and
corresponds to modify General Relativity, by modifying the
Einstein-Hilbert action. Modification of GR, subsequently, leads to
new formulation in gravity. The models in this class are$ f(R)$ and
$f(T)$ gravity \cite{DeFelice:2010aj,Sotiriou:2008rp,Nojiri:2010wj,
Nojiri:2006ri,deMartino:2015zsa,Beltran:2015hja,Basilakos:2015yoa,
KhodamMohammadi:2010py,Cai:2015emx,Iorio:2015rla}, scalar-tensor
theories \cite{Fujii:2003pa}, braneworld models
\cite{Rubakov:1983bb, Gogberashvili:1998iu, Gogberashvili:1999tb,
Maartens:2010ar} Gauss-Bonnet gravity \cite{Nojiri:2005vv,
Nojiri:2006je,Nojiri:2005jg, Nojiri:2005am} and so on.\\ Other
category is based on presence of an exotic component in stress
energy tensor, with sufficiently negative-pressure. This fluid which
is known as dark energy, accounts for roughly 75 percent of the
universe energy density today. Big variety of dark energy models are
proposed, nevertheless the nature and mechanism of dark energy is
not known yet. One of the most famous models, vastly used in
literature, is cold dark matter plus a cosmological constant named
($\Lambda$CDM) model. It explains the scenario of acceleration of
the universe and has an acceptable compatibility with recent
observational data \cite{Jassal:2006gf}, \cite{Wilson:2006tr},
\cite{Davis:2007na}, \cite{Allen:2007ue}. However this model suffers
from distinct problems; Fine tuning and coincidence. This made
theorists seek for some dynamical models instead
\cite{Copeland:2006wr, Nojiri:2010wj}. Actually any offered dark
energy model must entail all aspects of quantum theory, particle
physics and general relativity. One approach is holographic
principle, according to which, the entropy of a system scales not
with its volume but with its surface area \cite{Bekenstein:1973ur,
Ogushi:2004wf}. The motivation for this, was first arisen from
Bekenstein's entropy bound, $S\leq\pi M_p^2 L^2 $, from which it is
implied that in entropies well below this bound, quantum field
theory fails.\ Imposing a relation between UV and IR cut-offs, as
indicated in \cite{Cohen:1998zx}, conciliated this problem. This
relationship was established by using the limit set by black hole
formation, that is $L^3\Lambda^4\leq\pi M_p^2 L^2$ where $ M_p$ is
the reduced Planck mass. A holographic DE model where the IR
cut-off is given by the Ricci scalar and the Gauss-Bonnet (GB)
invariant was proposed in \cite{Saridakis:2017rdo}.\\ Recently, many
dynamical DE models, against rigid concordance model ($\Lambda$CDM),
has been proposed. The energy density of these models composed of
terms like $\dot{H},H\dot{H},H^{2}$ etc., which are studied in many
papers, e.g. \cite{Gomez-Valent:2014rxa, Gomez-Valent:2014fda} and
the role of terms like $H^{3}$, $\dot{H}H^{2}$ and $H^{4}$ in the
evolution of early universe has been investigated \cite{Lima:2012mu,
Perico:2013mna, Basilakos:2013xpa, Bleem:2014iim, Lima:1995ea}. It
is worthwhile to mention that in one form of these models
($\rho_D=C_0+C_1 H^2+C_2 \dot{H}$), authors concluded that their
model indicate a significantly better agreement with observations as
compared to the concordance $\Lambda$CDM model \cite{Sola:2015wwa}. \\
The studied model in present paper has been firstly proposed by
\cite{Granda:2013gka}, named natural scaling for DE, so that we
called it latter by Gauss-Bonnet DE model \cite{Karimkhani:2015gja}.
It complies the holographic principle and obeys the above bound for
black hole formation. The related energy density is proportional to
the Gauss-Bonnet 4-dimensional invariant, ${\cal G}$, in such a way
that it has the valid dimension of energy density
\cite{Granda:2013gka}. This invariant is used in corrections of low
energy string gravity. The GB-DE energy density is composed of
powers of Hubble parameter and its derivative. Many authors have
used the GB term in the bulk, coupled with some scalar fields or DE
models \cite{Saridakis:2007wx, BouhmadiLopez:2011xi,
Belkacemi:2011zk}. Also the reconstruction of the holographic DE in
the framework of the modified GB gravity was performed in
\cite{Jawad:2014kka} and other applications of the GB gravity in the
context of the holographic principle have been studied in
\cite{Zeng:2013mca, Li:2013cja, Andrade:2016rln}. Moreover the GB
term is employed in dark energy context with different forms in the
action, like coupled to some scalar field, used in modified theories
\cite{Nojiri:2005vv, Nojiri:2006je, Nojiri:2005jg, Nojiri:2005am},
or as modified dark energy models as in \cite{Karimkhani:2015gja,
Granda:2014zea}. Specially authors in \cite{Karimkhani:2015gja},
showed that only modified GB-DE has capability to have stability
against the density perturbation. The investigation of the cosmic
evolution and
the compatibility with the observations can help us to judge about GB-DE models.\\
This paper is organized as follows: In Sec.II, we review the
Gauss-Bonnet universe, in background and perturbations point of
views. In Sec.III, we proceed to data analysis and the results of
these methods for our models and at last finished our work with some
concluding remarks.
\section{ The Gauss-Bonnet universe}
\subsection{Background equations}
The energy density of GB-DE has been firstly introduced by
\cite{Granda:2014zea}
\begin{equation}\label{eq1}
\rho_{d}=\alpha\cal G,
\end{equation}
where $\alpha$ is a dimensionless parameter and ${\cal G}$ is the
4-dimensional Gauss-Bonnet invariant which is defined as
\begin{equation}\label{eq2}
\mathcal{G}=R^{2}-4R_{\mu\nu}R^{\mu\nu}+R_{\mu\nu\eta\gamma}R^{\mu\nu\eta\gamma}.
\end{equation}
It's easy to see that for the flat FRW background,
$ds^2=-dt^2+a(t)^2\sum^{3}_{i=1}(dx^i)^2$, the GB dark energy
density (\ref{eq1}) can be written as
\begin{equation}\label{eq3}
\rho_{d}=24\alpha\left(H^4+H^2\dot{H}\right).
\end{equation}

\subsection*{Modified Gauss-Bonnet Dark energy}
Modified GB-DE (MGB-DE) has the following energy density
\cite{Granda:2013gka}
\begin{equation}\label{eq4}
\rho_{d}=\gamma H^4+\beta H^2\dot{H},
\end{equation}
whit two independent free parameter $\gamma$ and $\beta$. For a
single component universe (in the absence of matter), the Friedmann
equation with the energy density given by (\ref{eq2}), in the flat
FRW background takes the form
\begin{equation}\label{eq5}
\beta\frac{dH}{dt}+\gamma H^2-\frac{3}{\kappa^2}=0,
\end{equation}
where $\kappa^2=8\pi G=M_p^{-2}$. With suitable initial condition, this equation is solved and discussed in\cite{Granda:2013gka}.\\
In presence of matter(baryonic and dark), the Friedmann equation
becomes non-linear and does not have exact solution. Adding the
matter term $\rho_m=(\rho_c+\rho_b)=\rho_{m0}a^{-3}$, the Friedmann
equation reads
\begin{equation}
a \tilde \beta E^3 E' - E^2 + \tilde\gamma E^4 +\frac{
\Omega_m}{a^3}=0\, \label{eqc1}
\end{equation}
where $\tilde{\gamma}=\kappa^2H_0^2\gamma/3$,
$\tilde{\beta}=\kappa^2H_0^2\beta/3$ and
$\Omega_{m0}=\kappa^2\rho_{m0}/(3H_0^2)$
($\Omega_m=\kappa^2\rho_m/(3H^2)$). The scaled Hubble parameter is
defined as $E=H/H_0$. With the initial condition, $E(1)=1$, and
different amounts of parameters, this equation could be solved
numerically.\\Since we are interested in late universe data or their
mixture with those of background solutions in recent time, we can
neglect radiation in the evolutionary equations.
\subsection*{The Interacting MGB model}
The interacting MGB (IMGB) model of DE is also introduced as a
second model. Dark energy models in GR, suffer from the coincidence
problem referred to energy density orders of dark matter and dark
energy. This problem could be solved by assuming continuous energy
exchange between dark sectors. The signature of non-gravitational
interaction term $\bar Q$, in the continuity equations, shows the
direction of energy transfer
\begin{eqnarray}
\dot{\bar{\rho}}_c+3H\bar{\rho}_c&=& \bar{Q},
\label{rhocb}\\
\dot{\bar{\rho}}_d+3H(1+w)\bar{\rho}_d&=-&\bar{Q}. \label{rhoxb}
\end{eqnarray}
Here $w=\bar{P}_d/\bar{\rho}_d$. The $\bar Q$ as the rate of energy
density transfer is usually introduced as
\begin{equation} \label{Q}
\bar Q=-\left(\Gamma_m\bar{\rho}_m+\Gamma_d\bar{\rho}_d \right) \,
\end{equation}
where $\Gamma_i$'s($\Gamma_m$ or $\Gamma_d$) are constant energy
density transfer rates and show the decay of dark matter to dark
energy, or vice versa (Baryons (b) and photons ($\gamma$) are not
coupled to dark energy). We are interested in the special case
$\Gamma_d=0$ and choose $\Gamma_m=3H\xi^2$. Hence from the
continuity equation, the dark matter density is
\begin{equation}
{\bar{\rho}}_m={\rho_0 }_m a^{-3(1-\xi^2)}
\end{equation}
and also the Eq. (\ref{eqc1}) changes to
\begin{equation}
a \tilde \beta E^3 E' - E^2 + \tilde\gamma E^4 +\frac{ \Omega_b}{a^3}+\frac{ \Omega_c}{a^{3(1-\xi^2)} }=0\;.\label{eqcint1}
\end{equation}
\subsection*{ MGB with a constant}
As a third model, we consider the MGB with an arbitrary constant
like the approach in \cite{Gomez-Valent:2014fda} and
\cite{Arab:2017gae}. For this, we added a constant $n_0$ to the Eq.
(\ref{eqc1}) directly. It is worth noting that for a very small
value of $H(z)$, it reduced to familiar $ \Lambda$LCDM model.
\begin{equation}
a \tilde \beta E^3 E' - E^2 + \tilde\gamma E^4 +\frac{ \Omega_m}{a^3}+n_0=0\;.\label{eqccc}
\end{equation}
\subsection{The linear perturbed equations}
In perturbation theory, we consider a perturbed spacetime that is
close to the background spacetime. This means that there exists a
coordinate system on the perturbed spacetime, where its metric can
be written as
\begin{equation}
g_{\mu\nu}=\bar{g}_{\mu\nu}+\delta g_{\mu\nu}.
\end{equation}
Here $\bar{g}_{\mu\nu}$ is the metric of the background. Thus metric
perturbations are divided into a scalar, vector and a tensor part,
which do not couple to each other in first-order perturbation theory
and evolve independently. Scalar perturbations are of special
importance. They couple to density and pressure perturbations and
cause gravitational instabilities. This make overdensities grow and
become more overdense. The outcome is formation and growth of Large
Scale Structure (LSS), from small initial perturbations. In order to
study the linear perturbation theory, we start with perturbation
equations. In the perturbed FRW universe, with scalar perturbations
and in absence of anisotropic stress, the line element is
\begin{equation}\label{eq:line-element}
ds^2=-(1+2\Phi)dt^2+a^2(t)(1-2\Psi)d\vec{x}^2\;,
\end{equation}
where $\Phi$ and $\Psi$ are metric perturbations known as the
Bardeen potentials. Perturbations in density (matter or energy) and
pressure are
\begin{equation}
\rho=\bar{\rho}+\delta \rho
\end{equation}
\begin{equation}
p=\bar{p}+\delta p
\end{equation}
where $\bar{p}$ and $\bar{\rho}$ are pressure and density of background. The perturbed energy momentum tensor is
\begin{equation}
T^\mu_{\;\;\nu}=\bar{T}^\mu_{\;\;\nu}+\delta T^\mu_{\;\;\nu}
\end{equation}
The DE component is expected to be smooth and we consider perturbations only on the matter
component of the cosmic fluid. The energy-momentum continuity
equation needs $ T^\mu_{\;\;\nu;\mu}=0$. In
absence of interaction between dark matter and dark energy and in Fourier space this equation leads to
\begin{equation}
\dot{\delta_{\rm m}}=(1+\omega_{\rm m})(3 \dot{\Psi}+\frac{k}{a}
\theta_{\rm m}),
\end{equation}
\begin{equation}
\dot{\theta_{\rm m}}+(1-3 \omega_{\rm m}) H{\theta_{\rm
m}}=\frac{k}{a} (\Phi+\frac{\omega_{\rm m}}{1+\omega_{\rm m}}
\delta_{\rm m}),
\end{equation}
in which $\delta_m(=\delta \rho_m/\rho_m)$ is dark matter density contrast and $\theta_{\rm m}$ is the
divergence of velocity field. We are interested in the case of non-relativistic fluid $ (\omega_{\rm m}=0)$
and scales much smaller than Hubble radius $(k\gg aH)$. So that the above equations result into a second
order differential equation, for evolution of matter density contrast. In terms of scale factor it reads
\begin{equation}
\delta^{\prime\prime}_{\rm m}+\left(\frac{3}{a}+\frac{E^{\prime}}{E}\right)\delta^{\prime}_{\rm m}-
\frac{3}{2a^2 E^2}\Omega_{\rm m}\delta_{\rm m}=0 \;.
\label{eqc2}
\end{equation}
For coupled MBG dark energy, the changes are exhibited in the background evolution equations, in $\Omega_{m}$ term.\\
Solving the system of equations (\ref{eqc1})
and (\ref{eqc2}) gives the evolution of density contrast for the models. In order to study structure formation and compare models with data, it is needed to use some definitions. The first concept is the growth rate function defined with the following equation
\begin{equation}
\label{faa}
f(a)=\frac{d\ln \delta_{m}}{d\ln a}\;.
\end{equation}\
The observable that we need to measure in structure formation
context, is $f\sigma_8$, in which, $\sigma_8$ is
\begin{equation}
\label{sig8} \sigma_{8}(a)=\sigma_{8,0}\frac{
\delta_{m}(a)}{\delta_{0}}\;
\end{equation}
where $\delta_{0}$ is the density contrast in $a=1$.\\
Another important quantity we can refer to is the $\gamma$-index.
This index is related to matter perturbations and is defined via
$f(z)\simeq \Omega_m (z)^{\gamma(z)}$, so the growth index
$\gamma(z)$ can be written as
\begin{equation}
\label{eq:gamma} \gamma(z)\cong \frac{\ln f(z)}{\ln\Omega_m (z)}
\end{equation}
\section{Observational constraints}
In this section we use data analyzing methods in order to find the
best fit values of the parameters in background and perturbed level
for MGB-DE universe. To study the expansion history and the growth
rate of structures, we ought to define some observables at first.
The most important are the background expansion indicators such as
distance modulus of Supernovae type Ia, Hubble parameter, Baryon
acoustic oscillations (BAO) and CMB power spectrum. The observable
related to perturbation growth rate of structures is $ f\sigma _8 $ data and is taken into account correspondingly. \\
The respective parameters to be defined are: parameters of the MGB-DE models, $ \tilde \beta$,
$ \tilde \gamma$,$\xi$,$n_0$ plus usual cosmological parameters like
current matter and baryon density parameters, $\Omega_{\rm m}^0$ , $\Omega_{\rm b}^0$ and $h=H_{0}/100$ (normalized Hubble constant).
\\ Available observational data sets, used for these calculations are: distance modulus
of Supernovae Type Ia, Baryon acoustic oscillations (BAO), Hubble evolution data, growth rate data f$\sigma 8$ and WMAP data for CMB which will be explained;
\subsection{observables}
The main evidence for cosmic accelerated expansion is Supernovae.
Measuring the luminosity distance of these objects not only gives
useful information about history of early universe but also
constrain model parameters in low and intermediate redshifts
confidently. Referred catalogue is the SnIa distance module from
Union 2.1 sample \cite{Suzuki:2011hu}, which includes 580 SnIa over
the redshift range
$0<z<1.4$.\\
By introducing covariant matrix $\mathbf{C}_{\rm sn}$, which
includes systematic uncertainties and correlation information of
SNIa data sets, from \cite{Suzuki:2011hu}, the $\chi^2$ for SnIa is
given by:
\begin{equation}\label{eq:xi2-sn}
\chi^2_{\rm SN}=\mathbf{U}^{T}\mathbf{C}_{\rm sn}^{-1}\mathbf{U}\;,
\end{equation}
in which
\begin{equation}
\mu_{\rm
th}(z)=5\log_{10}\left[(1+z)\int_0^z\frac{dx}{E(x)}\right]+\mu_0,
\end{equation}
and
\begin{equation}
\mathbf{U}=\mu_{\rm th}(z_i)-\mu_{\rm ob}(z_i)
\end{equation}
are the theoretical distance modulus and the difference matrix $\mathbf{U}$, accordingly.
Because of applying covariance matrix $\mathbf{C}_{\rm sn}$ we do not regard the noisy parameter $\mu_0$. \
Baryon acoustic oscillations (BAO), are the imprint of oscillations
in the baryon-photon plasma on the matter power spectrum. They are
less affected by nonlinear evolution so they can be used as a
standard ruler. The BAO data can be applied to measure the angular
diameter distance ${D_A}$ and the expansion rate of the Universe
${H(z)}$ either separately or through the combination. We utilize 6
reliable measurements of BAO indicator, including Sloan Digital Sky
Survey (SDSS) data release, 7 (DR7) , SDSS-III Baryon Oscillation
Spectroscopic Survey (BOSS) , WiggleZ survey and 6dFGS survey. BAO
observations contain measurements from redshift interval,
$(0.1<z<0.7)$, summarized in Table.I.
\begin{table}[h]
\centering
\begin{tabular}{llcccc}
\hline
\hline Redshift & Data Set & $r_s/D_V(z;\{\Theta_p\})$ & Ref.\\ \hline
0.10 & 6dFGS & $0.336\pm0.015$ & \cite{Beutler:2011hx} \\
0.35 & SDSS-DR7-rec & $0.113\pm0.002$& \cite{Padmanabhan:2012hf} \\
0.57 & SDSS-DR9-rec & $0.073\pm0.001$ & \cite{Anderson:2012sa} \\
0.44 & WiggleZ & $0.0916\pm0.0071$ & \cite{Blake:2011en} \\
0.60 & WiggleZ & $0.0726\pm0.0034$ & \cite{Blake:2011en} \\
0.73 & WiggleZ & $0.0592\pm0.0032$ & \cite{Blake:2011en} \\
\hline
\hline
\label{tab:bao}
\end{tabular}
\caption{\label{baodata} Observed data for BAO \cite{Hinshaw:2012aka}. }
\end{table}
The $\chi$ square for BAO, as mentioned in \cite{Hinshaw:2012aka}, is
\begin{equation}\label{eq:xi2-bao}
\chi^2_{\rm BAO}=\mathbf{Y}^{T}\mathbf{C}_{\rm BAO}^{-1}\mathbf{Y}\;,
\end{equation}
where $\mathbf{Y}=(d(0.1)-d_{1},\frac{1}{d(0.35)}-\frac{1}{d_2},\frac{1}{d(0.57)}-
\frac{1}{d_3},d(0.44)-d_{4},d(0.6)-d_{5},d(0.73)-d_{6})$ and
\begin{equation}\label{eq:d(z)}
d(z)=\frac{r_{\rm s}(z_{\rm drag})}{D_V(z)}\;,
\end{equation}
with
\begin{equation}\label{eq:com-sound-H}
r_{\rm s}(a)=\int_0^{a}\frac{c_{\rm s}da}{a^2H(a)}\;,
\end{equation}
where $r_{\rm s}(a)$ is the comoving sound horizon at the baryon drag epoch, $c_{\rm s}$ is the baryon sound speed and $D_V(z)$ is defined by:
\begin{equation}\label{eq:dv-bao}
D_V(z)=\left[(1+z)^{2}D^{2}_{\rm A}(z)\frac{z}{H(z)}\right]^{\frac{1}{3}}\;,
\end{equation}
that $D_{\rm A}(z)$ is the angular diameter distance.
We used the fitting formula for $z_{\rm d}$ from \cite{Eisenstein:1997ik} and the baryon sound speed is given by:
\begin{equation}\label{eq:bary-soun}
c_{\rm s}(a)=\frac{1}{\sqrt{3(1+\frac{3\Omega_b^0}{4\Omega_{\gamma}^0}a)}}\;,
\end{equation}
where we set $\Omega_{\gamma}^0=2.469\times 10^{-5} h^{-2}$
\cite{Hinshaw:2012aka}. The covariance matrix $\mathbf{C}_{\rm
BAO}^{-1}$ in Eq. (\ref{eq:xi2-bao}), was obtained by
\cite{Hinshaw:2012aka}
\begin{eqnarray}\label{eq:cij-bao}
\left(
\begin{array}{cccccc}
4444.4 & 0. & 0. & 0. & 0. & 0. \\
0. & 34.602 & 0. & 0. & 0. & 0. \\
0. & 0. & 20.6611 & 0. & 0. & 0. \\
0. & 0. & 0. & 24532.1 & -25137.7 & 12099.1 \\
0. & 0. & 0. & -25137.7 & 134598.4 & -64783.9 \\
0. & 0. & 0. & 12099.1 & -64783.9 & 128837.6
\end{array}
\right)\;.\nonumber
\end{eqnarray}\
The data related to cosmic microwave background, CMB, is used to
study early universe and dark energy models. CMB shift parameter, is
associated with the location of the first peak $\mathbf{L}_{\rm
1}^{TT}$ of the CMB temperature perturbation spectrum. It provides a
useful data to constrain dark energy models. The position of this
peak is given by $(l_{\rm a},R,z_{\ast})$, where $R$ is the scale
distance to recombination and is given for spatially flat cosmology
\begin{equation}
{R} = \sqrt{\Omega_{\rm m}^{0}}H_{\rm 0}D_{\rm A}(z_{\ast})\;.
\end{equation}\label{eq:R-cmb}
The quantity $l_{\rm a}$ is given by
\begin{equation}
{l_a} = \pi\frac{D_{\rm A}(z_{\ast})}{r_s(z_{\ast})},
\end{equation}
and $r_{\rm s}(z)$ is the comoving sound horizon which is defined in
Eq.~(\ref{eq:com-sound-H}). The fitted formula for $z_{\ast}$ , the
redshift of decoupling, is given in \cite{Hu:1995en}. For the WMAP
data set we have \cite{Hinshaw:2012aka}
\begin{equation}\label{eq:x-cmb}
\mathbf{X}_{\rm CMB}= \left(
\begin{array}{c}
l_{\rm a}-302.40 \\
R-1.7264 \\
z_{\ast}-1090.88
\end{array}
\right).
\end{equation}
By defining the inverse matrix
\begin{eqnarray}
\mathbf{C}_{\rm CMB}^{-1}=\left(
\begin{array}{ccc}
3.182 & 18.253 & -1.429 \\
18.253& 11887.879& -193.808\\
-1.429& -193.808& 4.556
\end{array}
\right),
\end{eqnarray}
the $\chi^2_{\rm CMB}$ is obtained by:
\begin{equation}\label{eq:xi2-cmb}
\chi^2_{\rm CMB}=\mathbf{X}_{\rm CMB}^{T}\mathbf{C}_{\rm CMB}^{-1}\mathbf{X}_{\rm CMB}\;.
\end{equation}\
The observed H(Z) data, are used to constrain cosmological
parameters. The advantage of using OHD is that they are acquired
directly from model-independent observations. Generally Hubble
parameter measurements are based on galaxy differential age and
radial BAO size methods. To avoid correlations in our calculations,
we use a Hubble data catalogue that is independent to BAO
measurements and includes 30 data points in the range of $0\leqslant
z \leqslant1.96$, as used in \cite{Sola:2016jky}. The $\chi^2$ for
this data set is:
\begin{equation}\label{eq:xi2-H}
\chi^2_{\rm H}=\sum_i\frac{[H(z_i)-H_{\rm ob,i}]^2}{\sigma_i^2}\;.
\end{equation}
\\The last data we refer to, is the growth rate data which probes structure formation
on large scales. The imprint of dark energy on structure formation,
made it an efficient tool for debating on dark energy models
\cite{peebles1993principles}. The $f\sigma_8(z)$ data were derived
from redshift space distortions, from galaxy surveys including PSCs,
2DF, VVDS, SDSS, 6dF, 2MASS, BOSS and WiggleZ. The data with their
references are shown in Table.~\ref{tab:fsigma8data}. The
$\chi^2_{\rm f\sigma_8}$ is written as
\begin{equation}\label{eq:xi2-fs}
\chi^2_{\rm f\sigma_8}=\sum_i\frac{[f\sigma_8(z_i)-f\sigma_{8,\rm ob}]^2}{\sigma_i^2}\;.
\end{equation}
\begin{table}[!h]
\caption{The $f\sigma_8(z)$ growth data.} \tiny{
\begin{tabular}{| c | c | c | }
\hline
\hline
z & $f\sigma_8(z)$ & Ref. \\
\hline
$0.02$ & $0.360\pm0.040$ & \cite{Hudson:2012gt}\\
$0.067$ & $0.423\pm0.055$ & \cite{Beutler:2012px}\\
$0.10$ & $0.370\pm0.130$ & \cite{Feix:2015dla}\\
$0.17$ & $0.510\pm0.060$ & \cite{Percival:2004fs}\\
$0.35$ & $0.440\pm0.050$ & \cite{Song:2008qt,Tegmark:2006az}\\
$0.77$ & $0.490\pm0.180$ & \cite{Guzzo:2008ac,Song:2008qt}\\
$0.25$ & $0.351\pm0.058$ & \cite{Samushia:2011cs}\\
$0.37$ & $0.460\pm0.038$ & \cite{Samushia:2011cs}\\
$0.22$ & $0.420\pm0.070$ & \cite{Blake:2011rj}\\
$0.41$ & $0.450\pm0.040$ & \cite{Blake:2011rj}\\
$0.60$ & $0.430\pm0.040$ & \cite{Blake:2011rj}\\
$0.60$ & $0.433\pm0.067$ & \cite{Tojeiro:2012rp}\\
$0.78$ & $0.380\pm0.040$ & \cite{Blake:2011rj}\\
$0.57$ & $0.427\pm0.066$ & \cite{Reid:2012sw}\\
$0.30$ & $0.407\pm0.055$ & \cite{Tojeiro:2012rp}\\
$0.40$ & $0.419\pm0.041$ & \cite{Tojeiro:2012rp}\\
$0.50$ & $0.427\pm0.043$ & \cite{Tojeiro:2012rp}\\
$0.80$ & $0.470\pm0.080$ & \cite{delaTorre:2013rpa}\\
\hline
\hline
\end{tabular}}
\label{tab:fsigma8data}
\end{table}
\subsection{Analysis}
We have proceeded joint data sets, consisting of cosmological
data, in order to study the models. Depending on model, there are three groups of free parameters in our analysis;
$p_1=\{h,\Omega_{\rm m},\Omega_{\rm
b},\tilde{\beta},\tilde{\gamma}\}$, $p_2=\{h,\Omega_{\rm
m},\Omega_{\rm b},\tilde{\beta},\tilde{\gamma},\xi\}$,
$p_3=\{h,\Omega_{\rm m},\Omega_{\rm
b},\tilde{\beta},\tilde{\gamma},n_0\}$. Datasets are selected in a way that we can study the models in
late and early universe by mixture or pure high and low redshift data. We have found the best value of the parameters and calculated chi-square $\chi^2_{\rm tot}$ for joint datasets. The performance of a model could be tested
via the Aakaike statistical information criterion AIC. It accounts
the number of degrees of freedom and the number of fitting
parameters.
\begin{equation}\label{eq:AIC}
{\rm AIC}=\chi^2_{min}+2n_{\rm fit}.
\end{equation}
To test the effectiveness of models $M_i$ and $M_j$, one considers
the difference amount $\Delta AIC_{ij}=\vert AIC_i-AIC_j\vert$. The
larger the value of $|\Delta AIC_{ij}|$, the higher the evidence
against the model with higher value of AIC. The range $2 \leq|\Delta
AIC_{ij}|\leq 6$, indicating a positive such evidence and for
$|AIC_{ij}|\geq 6$ a significant such evidence is concluded. Usually
one of these models is the rigid $\Lambda$CDM model which has a good
consistency with cosmological observations.

\section{Discussion and results}
The first joint data set used in this paper is, Hubbe+SNIa+f$\sigma_8$+CMB+BAO. Total $\chi^2$ for this set is written as:
\begin{equation}
\chi^2_{\rm tot1}=\chi^2_{\rm Hubble}+\chi^2_{\rm
f\sigma_8}+\chi^2_{\rm SN}+\chi^2_{\rm BAO}+\chi^2_{\rm CMB}.
\end{equation}
The results of constraint of free parameters are classified based on
MGB models in Table \ref{tab:analys1}. The calculated $\Delta AIC$
amounts referring to the related amount of $\Lambda$CDM (
$\chi^2_{\rm \Lambda CDM}$=575.205). It shows that there is not any
significant evidence against or in favor of $\Lambda$CDM for all
models, since $\Delta AIC<6$. However, in comparison between models,
no one has a significant difference with others.
\begin{table}[!h]
\caption{The best value parameters and their 1-$\sigma$ uncertainty
for the MGB models with joint dataset1 ($Hubbe+SNIa+ f\sigma_8+CMB+BAO$).}
\
\begin{tabular}{| c | c | c | c |}
\hline \hline
parameter &$MGB$&$ IMGB$&$MGB+n_0$\\
\hline
\hline
$h$ &$0.711978^{+0.003873}_{-0.003798}$&$0.710614^{+0.003715}_{-0.003670}$ &$0.711175^{+0.003860}_{-0.003797}$\\
\hline
$\Omega_{\rm m}^0$&$0.212647^{+0.004686}_{-0.004415}$& $0.212519^{+0.004564}_{-0.004423}$&$0.213151^{+0.003745}_{-0.003638}$\\
\hline
$\Omega_{\rm b}^0$&$0.044142^{+0.000498}_{-0.000494}$&$0.044335^{+0.000524}_{-0.000495}$&$0.0442440^{+0.000513}_{-0.000507}$\\
\hline
$\tilde{\beta}$&$0.645516^{+0.000451}_{-0.000452}$&$0.642940^{+0.000441}_{-0.000442}$&$0.525705^{+0.000379}_{-0.000380}$\\
\hline
$\tilde{\gamma}$&$0.915838^{+0.000555}_{-0.000556}$&$0.924112^{+0.000537}_{-0.000536}$&$0.747644^{+0.000477}_{-0.000476}$\\
\hline
$\xi$&---&$0.303848^{+0.007741}_{-0.007966}$&---\\
\hline
$n_0$&---&---&$0.136901^{+0.003611}_{-0.003790}$\\
\hline
$\chi^2_{min}$&574.795&573.204&574.676\\
\hline
$\Delta AIC$&3.590&3.999&5.471\\
\hline \hline
\end{tabular}
\label{tab:analys1}
\end{table}\\
\begin{figure}[!h]
\includegraphics[scale=.87]{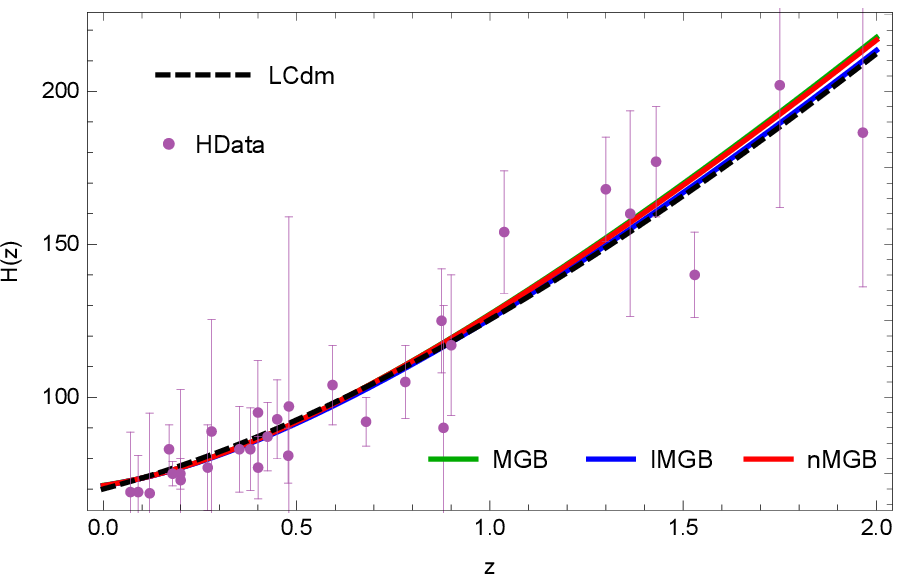}
\includegraphics[scale=.87]{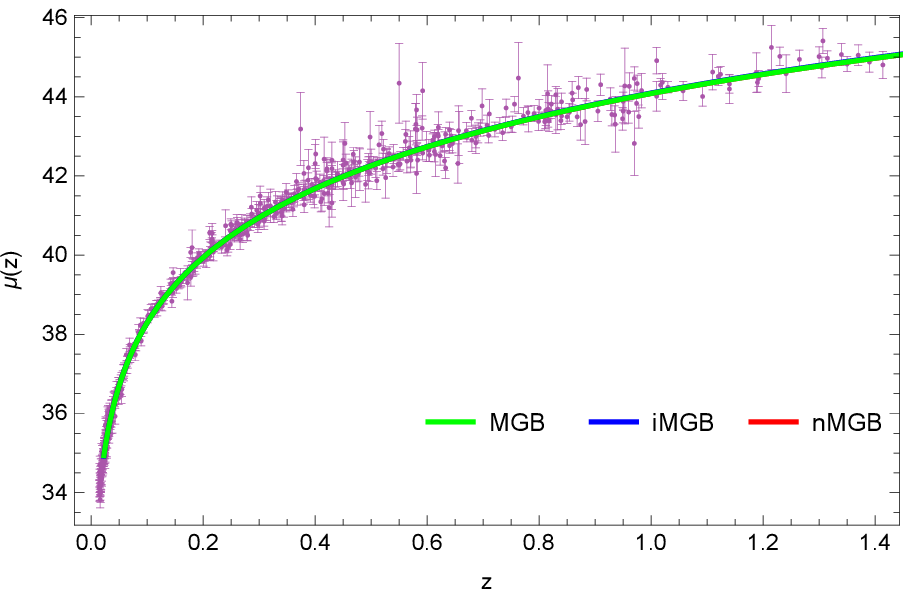}
\caption{\small{Hubble parameter and luminosity distance of for MGB
models with best values from data set ($Hubbe+SNIa+
f\sigma_8+CMB+BAO$). Observed data are indicated with error bars.}}
\label{fig1}
\end{figure}
In order to investigate the cases phenomenologically, we use the best values of parameters and study the main aspects of
the models.
\begin{figure}[!h]
\includegraphics[scale=.87]{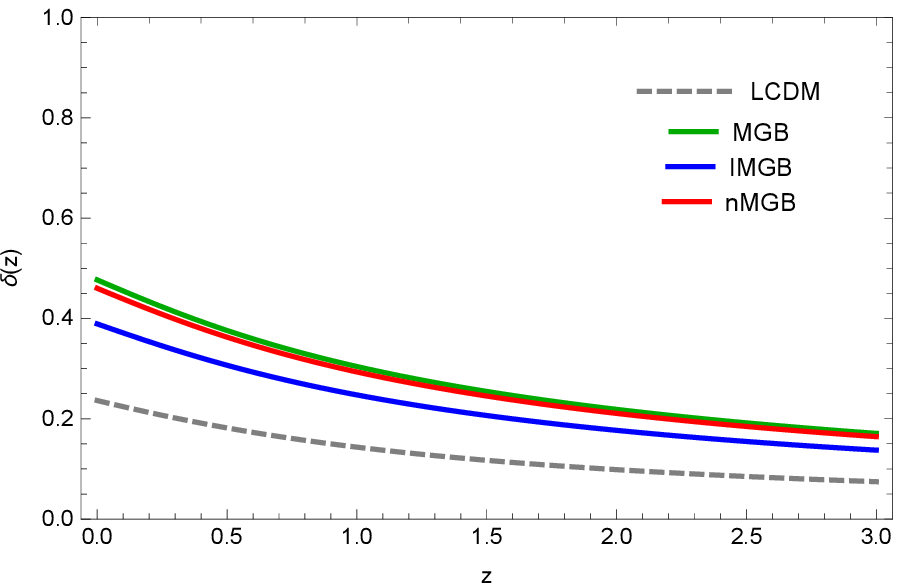}
\includegraphics[scale=.86]{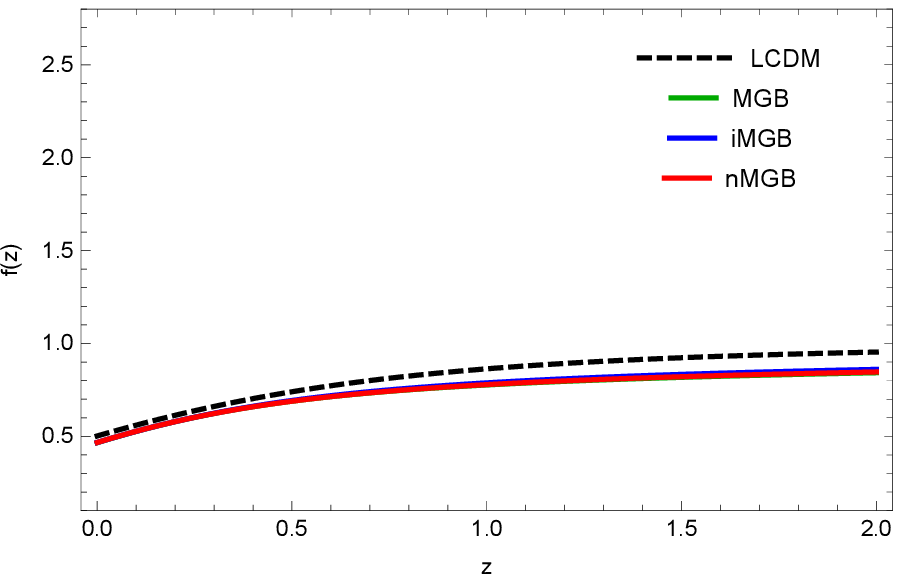}
\caption{\small{The density contrast $ \delta$, (left) and the growth rate function (right)
 for MGB like models with the best fit parameters from data set ($Hubbe+SNIa+ f\sigma_8+CMB+BAO$).}}
\label{fig2}
\end{figure}\\
In Fig.\ref{fig1}, the Hubble parameters of models, are shown and
compared with the data. They show acceptable treatments and explain
the evolution of universe properly.\ In the right panel, the
distance modulus of models are shown. Comparing the models with the
Union data, we see that plots are clearly well fitted to the data
owing to the large number of SNIa data in the constraining
process.\\To justify dark energy or modified gravity models, we
should study them in the structure formation process. Theories with
better predictions in this subject seem to be worthy to research
about. In Fig.\ref{fig2}, the density contrast and growth rate
function are plotted for the models with best fit parameters from
dataset ($Hubbe+SNIa+ f\sigma_8+CMB+BAO$). The density contrast for
IMGB model shows better competency with $\Lambda$CDM. In
Fig.\ref{fig3}, the $ f\sigma_8$ plots are shown. All MGB models
show very close treatments. They are near to $\Lambda$CDM and pass
through the data. In the right panel, $\gamma$ indices for MGB like
models are exhibited. The departures from $\Lambda$CDM index are
between 2-3 percents. This may have some reasons like present
experimental limits that may be alleviated by increasing the
accuracy of observations.
\begin{figure}[!h]
\includegraphics[scale=.86]{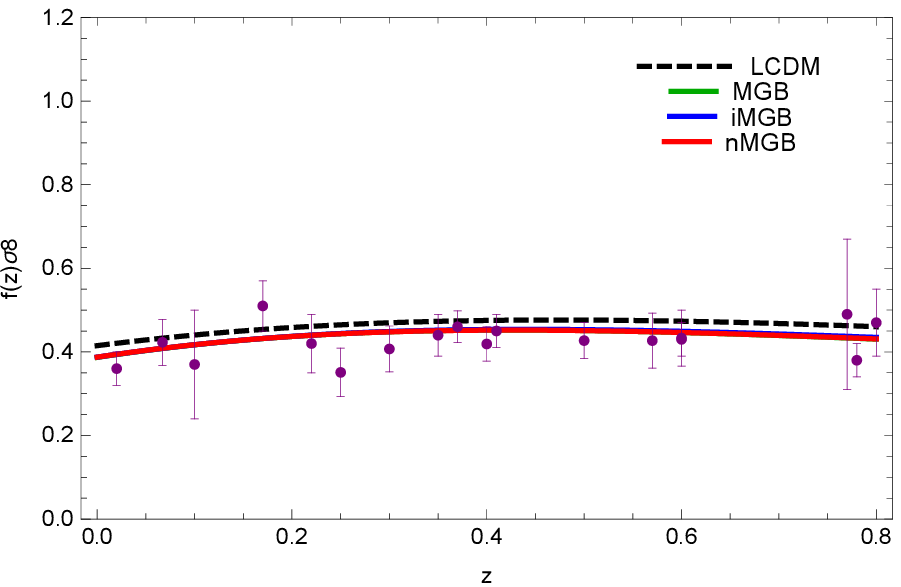}
\includegraphics[scale=.86]{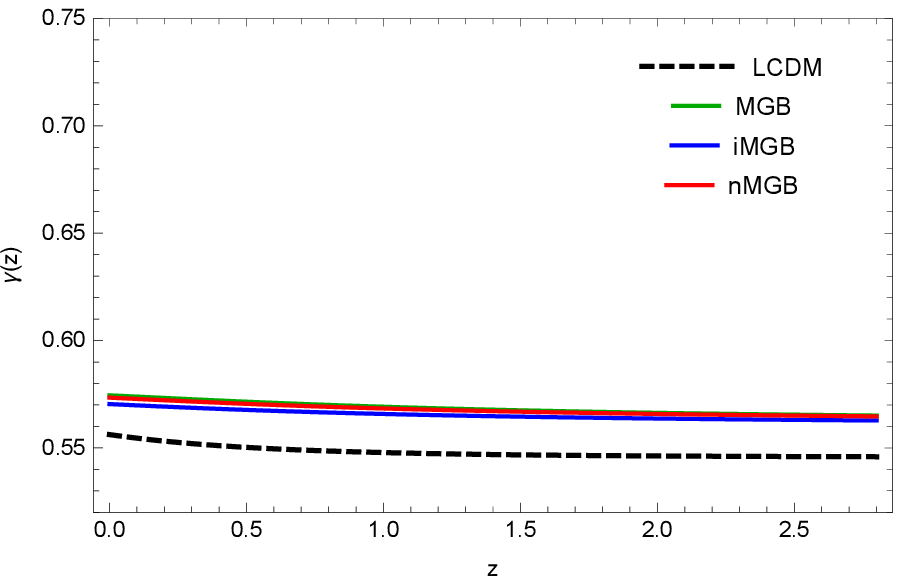}
\caption{\small{The $ f\sigma_8$ plot for MGB like models, LCDM and
observed data(left). The $ \gamma$ index plot for MGB like models
and LCDM(right) }} \label{fig3}
\end{figure}\\
Generally, variety of MGB-DE models predicts the evolution of
universe and structures in successful way and statistical results
are satisfactory. To clarify, two more datasets are
employed for late and early universe. The $\chi^2$ for the sets are: \\
data set 2 (late time):
\begin{equation}
\chi^2_{\rm tot2}=\chi^2_{\rm Hubble}+\chi^2_{\rm
f\sigma_8}+\chi^2_{\rm SN},
\end{equation}
and data set 3 (early time):
\begin{equation}
\chi^2_{\rm tot3}=\chi^2_{\rm SN}+\chi^2_{\rm BAO}+\chi^2_{\rm CMB}.
\end{equation}
The results of the above combinations are summarized in Table IV.
Statistically, In late universe, there is not any significant
difference between mentioned models of MGB. In early universe, due
to $|\Delta AIC|>6$, there is a ``strong'' evidence against MGB
models in favor of $\Lambda$CDM. The deduction is that, although MGB
models have enough performance in late universe, it suffers to some
problems at early universe. However at early time, MGB+$n_0$ model
is remarkably better than other two.
\begin{table}[!h]
\centering
\begin{tabular}{llc c c c}
\hline
\hline set/model & MGB & IMG & MGB+$n_0$ \\ \hline
$\Delta$AIC$_2$ & 0.855&1.365 &2.586 \\
$\Delta$AIC$_3$ & 14.09& 12.27 & 6.80 \\
\hline
\hline
\label{tab:aic}
\end{tabular}
\caption{The comparison between AIC of models}
\end{table}
\subsection{Conclusion}
We have studied Modified Gauss Bonnet dark energy with main
cosmological data sets. Applying the best obtained parameters to
study the model, showed that all versions of MGB-DE predicts the
expansion history and evolution of structures appropriately as well
as $\Lambda$CDM. If we use pure late universe data set, we see that
all versions of MGB-DE are successful in recent epoch, and there is
not any significant evidence against or in favor of $\Lambda$CDM,
whereas for early universe, statistical results indicate a
significantly better agreement for $\Lambda$CDM as compared to all
versions of MGB-DE models.\\
Observable show near treatments for the versions of MGB-DE. They are
highly sensitive to Hubble parameter as it is predictable. The
choice of data sets has a considerable effect on the outcome. Dark
energy perturbations that can impress the late time expansion of the
universe and evolution of structures, are ignored in this work. This
case can be investigated separately.

\acknowledgments

 We would like to express sincere gratitude to Dr. Ahmad Mehrabi for constructive
 comments and discussion.

\bibliography{reff}
\end{document}